\documentclass[aps,prd,onecolumn,eqsecnum,amsmath,nofootinbib,preprintnumbers]{revtex4}%

\usepackage[dvips]{color,graphicx}
\usepackage{amsfonts,amssymb,theorem,mathrsfs,times}
\usepackage{bm}
\usepackage[T1]{fontenc} % if needed
\usepackage{comment}
\usepackage{soul}
\usepackage[textsize=footnotesize,textwidth=2.cm]{todonotes}
\usepackage{float}

\newcommand{\ba}{\begin{array}}
\newcommand{\ea}{\end{array}}
\newcommand{\bi}{\begin{itemize}}
\newcommand{\ei}{\end{itemize}}
\newcommand{\bea}{\begin{eqnarray}}
\newcommand{\eea}{\end{eqnarray}}
\newcommand{\be}{\begin{equation}}
\newcommand{\ee}{\end{equation}}

\newcommand{\la}{\langle}
\newcommand{\ra}{\rangle}
\newcommand{\mr}{\mathrm}
\newcommand{\cd}{\cdot}
\newcommand{\cds}{\cdots}
\newcommand{\w}{\wedge}

\newcommand{\p}{\partial}

{\theorembodyfont{\upshape}
}
{\theorembodyfont{\upshape}
}
{\theorembodyfont{\upshape}
}
{\theorembodyfont{\upshape}
}
{\theorembodyfont{\upshape}
}
{\theorembodyfont{\upshape}
}

\newcommand{\dalm}{\kern1pt\vbox{\hrule height 0.9pt\hbox{\vrule width
0.9pt\hskip 2.5pt\vbox{\vskip 5.5pt}\hskip 3pt\vrule width 0.3pt}\hrule height
0.3pt}\kern1pt}

\begin{document}
%\thispagestyle{empty}
%\preprint{\hfill {\small {ICTS-USTC-14-12}}}
%<<<<<<<<<<<<< TITLE >>>>>>>>>>>>>>>%
\title{Gravitational waves in gauge theory gravity with a negative cosmological constant
}

%<<<<<<<<<<<<< AUTHOR >>>>>>>>>>>>>>>%
%\author{$^b$}
%
%\email{}

\author{Jianfei Xu\footnote[0]{e-mail address:
jfxu@seu.edu.cn}}

%<<<<<<<<<<<<< ADDRESS >>>>>>>>>>>>>>>%

\affiliation{
Shing-Tung Yau Center and school of mathematics, Southeast University, Nanjing, 210000, China}

%<<<<<<<<<<<<< DATE >>>>>>>>>>>>>>>%
\date{\today}

%======================================%
%<<<<<<<<<<<<< ABSTRACT >>>>>>>>>>>>>>>%
%======================================%
\begin{abstract}
In this paper, we discuss the gravitational waves in the context of gauge theory gravity with a negative cosmological constant. The gauge theory gravity is a gravity theory under gauge formulation in the language of geometric algebra. In contrast to general relativity, the background spacetime in gauge theory gravity is flat, the gauge freedom comes from the fact that equations in terms of physical quantities should be kept in a covariant form under spacetime displacement and rotation. Similar to the electromagnetism, the gauge formulation enables us to interpret the gravitational force as a gauge force on the background flat spacetime. The dynamical fields that describe the gravitational interactions are those position and rotation gauge fields introduced as the requirement of the gauge covariance. The gravitational field equations can be derived from the least action principle with the action as a gauge invariant quantity built from the covariant field strength. We discuss the gravitational wave solutions of the field equations with a negative cosmological constant, and show that these solutions are of Petrov type-N. We also discuss the velocity memory effect by calculating the velocity change of an initially free falling massive particle due to the presence of the gravitational waves.
\end{abstract}

%<<<<<<<<<<<<< PACS NUMBER >>>>>>>>>>>>>>>%
%\pacs{
%04.20.Cv, %Fundamental problems and general formalism
%04.20.Ha,%Asymptotic structure
%04.50.+h. %Gravity in more than four dimensions, Kaluza-Klein theory,
%     %unified field theories; alternative theories of gravity
%}

\maketitle

%======================================%
%<<<<<<<<<<<< SECTION I  >>>>>>>>>>>>>>%
%======================================%

\section{Introduction}
Since Einstein first predicted the existence of gravitational waves in the context of general relativity in 1916, the considerations on such kind of spacetime wave have never stopped. It was believed that a accelerating massive object will create gravitational waves much like a accelerating charged particle emits electromagnetic radiations. The first circumstantial evidence of the gravitational wave comes from the observation of the orbital decay of a binary neutron star system~\cite{Taylor1982}. The two neutron stars in this system is orbitally accelerating to each other, so the energy is carried away by the gravitational radiation, and this cause the orbit to decay. The direct evidence of the gravitational waves was presented by LIGO Science Collaboration and Virgo Collaboration~\cite{Abbott:2016blz, Abbott:2016nmj, Abbott:2017vtc, Abbott:2017oio, TheLIGOScientific:2017qsa}. These observations of the gravitational waves come from binary black hole and binary neutron star mergers. Compare to the light, the gravitational waves can travel unhindered among celestial bodies, so since then people have opened a new window to observe the universe.

The gravitational waves can carry more information of the physics in the vicinity of stars with strong gravitational fields, and thus can be used to test different kinds of gravity theories. One way to extract this information is through the so called memory effect. The original work on the memory effect is done by Zel'dovich and Polnarev~\cite{Zeldovich:1974gvh}, who claim that  freely falling detectors originally at relative rest will be displaced after the passing of a burst of gravitational radiation while with vanishingly small relative velocity. As a contrast, the velocity memory effect, which tells that after the passing of a burst of a gravitational wave, the initially at relative rest particles will deviate from each other with a nonvanishing constant velocity, is also proposed in~\cite{Braginsky1985}, and later studied in~\cite{Braginsky1987, Grishchuk:1989qa, Lasenby2017}. In~\cite{Braginsky1985}, the velocity memory effect is considered as a test particle moving in a weak gravitational wave at linear level. The nonlinear generalizations are later presented in~\cite{Christodoulou:1991cr, Thorne:1992sdb, Blanchet:1992br, Harte:2012jg}. The memory effects in constant curvature background in general relativity have been studied in dS~\cite{Hamada:2017gdg, Bieri:2017vni} and AdS~\cite{Chu:2019ssw}. The memory effect is also related to the soft theorem and black holes~\cite{Hawking:2016msc, Zhang:2017geq}. To the non-perturbative region, the exact gravitational plane wave solutions of the Einstein's vacuum equations have been invoked to discuss the velocity memory effect in~\cite{Zhang:2017rno, Zhang:2018srn}, where such effect might have some relevance for detecting gravitational waves.

In this paper, we discuss the gravitational waves in the context of gauge theory gravity with a negative cosmological constant. The possible existence of of a negative cosmological constant with a fundamental string-inspired motivation has been explored in~\cite{Visinelli:2019qqu}, and the consistency between their results and the observation data also indicate that the gravitational wave solution discussed here is also well allowed by observations. In general relativity, the gravitational waves in vacuum spacetime with cosmological constant have been classified and studied in~\cite{Bicak:1999ha, Bicak:1999hb}. The gauge theory gravity is a gravity theory under gauge formulation~\cite{Lasenby:1998yq} in the language of geometric algebra~\cite{Doran2003}. In contrast to general relativity where the dynamical metric field also works as its background, the background spacetime in gauge theory gravity is flat. The gauge freedom comes from the fact that equations in terms of physical quantities should be kept in a covariant form under spacetime displacement and rotation. Similar to the electromagnetism, the gauge formulation enable us to interpret the gravitational force as a gauge force on the background flat spacetime. There is no such kind of curved spacetime concept like in general relativity. Instead, we have two kinds of gauge fields living on the flat spacetime, one is the position gauge field and the other is the rotation gauge field, as dynamical fields to describe the gravitational interactions. The field equations of gravity can be derived from the least action principle where the action is built from the field strength by requiring the gauge invariance. In~\cite{Lasenby:2019gmi}, Lasenby discusses the black hole and gravitational wave solutions to the field equations in the gauge theory gravity without cosmological constant as well as the velocity memory effect in this case. In the present paper, we consider a set of Petrov type-N solutions satisfying the field equations with a negative cosmological constant. Among them, the gravitational wave solutions now can be written as an explicit expression of the position gauge field contains a spacetime function similar to that in the Siklos spacetime~\cite{Siklos1985, Podolsky:1997ni, Podolsky:1997ik}, which is the only non-trivial Einstein space conformal to non-flat pp waves in generality relativity. The impulsive gravitational waves with cosmological constant have been obtained in~\cite{Hotta:1992qy} and analysed more detail in~\cite{Podolsky:1997ri}. We choose the gravitational wave as an impulsive one, and study the velocity memory effect. We find that an initially free falling massive particle will gains a constant velocity change after the impulsive wave passed over.

The organization of this paper is as following. In section \ref{sec2}, we introduce the mathematical framework of geometric algebra with necessary notations and conventions. In section \ref{sec3}, the gauge theory gravity based on the geometric algebra is discussed, where we review the gauge formulation of this theory and derive the gravitational field equations. In section \ref{sec4}, we consider the gravitational wave solutions to the field equations with a negative cosmological constant. This solution can be shown to be type-N under the Petrov classification. In section \ref{sec5}, we discuss the velocity memory effect by calculating the velocity change of a massive particle after the passing of a gravitational impulsive wave.

%======================================%
%<<<<<<<<<<<< SECTION II >>>>>>>>>>>>>>%
%======================================%
\section{Geometric Algebra}\label{sec2}
People in ancient Greeks used Euclidean geometry to describe the world. With the developments of mathematics, different kinds of geometries had been discovered. Each of these new geometries has distinct algebraic properties. The mathematicians in nineteenth century paid a lot of attentions to place these geometries within a unified algebraic framework. The key insight of this process was made by W. K. Clifford, who attempted to unify Hamilton's quaternion and Grassmann's extensive algebra into a single mathematical system. Along this path, Clifford found a simple unified algebraic framework in which the inner and outer product of vectors are combined together, this was later called Clifford algebra. However, these achievements in mathematics did not successfully adopted in physics until D. Hestenes realised that the Clifford algebra is a better language in describing Dirac equation and quantum mechanics. In modern physics, there are a lot of algebraic systems employed, like vector analysis, Lie algebra, spinor calculus, differential forms and so on. D. Hestenes spent a lot of efforts on developing Clifford algebra into a complete language for modern physics, which he calls geometric algebra.

The geometric algebra is a mathematical tool of universal applicability. It provides a unified language for many of modern physics which are originally based on different kinds of mathematics. In the following subsections, we review the axiomatic development of the geometric algebra and introduce the conventions we adopted.

\subsection{Axiomatic development of geometric algebra}
The starting point is the vector space from which the entire algebra will be generated. The main axioms govern the properties of the geometric product for vectors are the following:
\bi
\item[(1)]Associative:
\be
a(bc)=(ab)c=abc\,.
\ee
\item[(2)]Distributive over addition:
\be
a(b+c)=ab+ac,~~~~(b+c)a=ba+ca\,.
\ee
\item[(3)]The square of any vector is a real scalar:
\be
a^2\in\mathcal{R}\,.
\ee
\ei
Here we use lowercase $a, b, \cds$ to denote vectors. By successively multiplying together vectors we can generate the complete algebra. The elements of this algebra are called multivectors, which are linear combinations of geometric products of vectors,
\be
A=\alpha(abc\cds)+\beta(ef\cds)+\cds\,,
\ee
where $\alpha, \beta$ are real scalars, and we use uppercase $A, B, \cds$ to denote multivectors. The geometric product can also be applied to multivectors and this product inherits the properties for vectors,
\bea
&&A(BC)=(AB)C=ABC\,,\\
&&A(B+C)=AB+AC,~~~~(B+C)A=BA+CA\,.
\eea
The vectors are usually called grade-1 multivectors, and any linear combination for vectors is still a grade-1 multivector.

Define the inner product for vectors by
\be\label{cd}
a\cd b=\frac{1}{2}(ab+ba)\,,
\ee
which is a real scalar by axioms, and the remaining antisymmetric part of the geometric $ab$ is define as the outer product for vectors,
\be\label{w}
a\w b=\frac{1}{2}(ab-ba)\,.
\ee
With these definitions, the geometric product for two vectors can be written as,
\be
ab=a\cd b+a\w b\,.
\ee

The outer product for vectors $a_1, a_2, \cds, a_r$ is denoted by $a_1\w a_2\w\cds\w a_r$ and this is a $\textbf{grade-r}$ multivector defined as following,
\be\label{ww}
a_1\w a_2\w\cds\w a_r=\frac{1}{r!}\sum(-1)^{\epsilon}a_{k_1}a_{k_2}\cds a_{k_r}\,,
\ee
where the sum runs over every permutation of the indices $k_1, \cds, k_r$ of $1, \cds, r$ with coefficient $+1$ for even or $-1$ for odd. Any multivector that can be written purely as the outer product of a set of vectors, thus with fixed grade is called a $\textbf{blade}$. Multivectors contains terms of one grade are called $\textbf{homogeneous}$. So a homogeneous multivector can be written as a sum of blades, and each blade is a geometric anticommuting vectors. An arbitrary multivector $A$ can be decomposed into a sum of homogeneous terms,
\be
A=\la A\ra_0+\la A\ra_1+\cds=\sum_r\la A\ra_r\,,
\ee
where the operator $\la~\ra_r$ projects onto the grade-r terms in the argument. The abbreviation $\la~\ra=\la~\ra_0$ is usually used to label the scalar part. Using the equations above, the geometric product of a vector and a grade-r multivector $A_r$ can be find,
\be
aA_r=a\cd A_r+a\w A_r\,,
\ee
where
\bea
a\cd A_r&=&\la aA_r\ra_{r-1}=\frac{1}{2}\left(aA_r-(-1)^rA_ra\right)\,,\\
a\w A_r&=&\la aA_r\ra_{r+1}=\frac{1}{2}\left(aA_r+(-1)^rA_ra\right)\,.
\eea
The inner product by a vector will lower the grade of a multivector by 1 while the outer product by a vector will increase the grade by 1. The geometric product of two homogenous multivectors can be decomposed as
\be
A_rB_s=\la A_rB_s\ra_{|r-s|}+\la A_rB_s\ra_{|r-s|+2}+\cds+\la A_rB_s\ra_{r+s}\,,
\ee
where the most general case for dot and wedge product can be defined,
\bea
A_r\cd B_s&=&\la A_rB_s\ra_{|r-s|}\,,\\
A_r\w B_s&=&\la A_rB_s\ra_{r+s}\,.
\eea
By recursively using \eqref{cd} and \eqref{ww}, one can show the following useful expressions,
\bea
a\cd(a_1\w a_2)&=&(a\cd a_1)a_2-(a\cd a_2)a_1\,,\label{useful1}\\
a\cd(a_1\w A_{r-1})&=&(a\cd a_1)A_{r-1}-a_1\w(a\cd A_{r-1})\,,~~r\ge3\label{useful1A}\\
a\cd(a_1\w a_2\w a_3)&=&(a\cd a_1)(a_2\w a_3)-(a\cd a_2)(a_1\w a_3)+(a\cd a_3)(a_1\w a_2)\,,\label{useful2}\\
(a_1\w a_2)\cd(b_1\w b_2)&=&(a_1\cd b_2)(a_2\cd b_1)-(a_1\cd b_1)(a_2\cd b_2)\,.\label{useful3}
\eea
In a mixed product among inner, outer and geometric, the inner products are performed before outer products, and both are performed before geometric product. So in some cases, we can omit the brackets without ambiguity.

\subsection{Frame and reciprocal frame}
Any $n$ dimensional vector space can be expended by a set of $n$ independent frame vectors. We use the symbols $e_1, \cds, e_n$ to denote the frame of a $n$-dimensional vector space. Based on these vectors, the entire geometric algebra $\mathcal{G}_n$ can be formed. The ingredients are scalar 1, vectors $e_i$, bivectors $e_i\w e_j$, trivectors $e_i\w e_j\w e_k$, and so on, up to the volume element,
\be\label{vol}
E_n=e_1\w e_2\w\cds\w e_n\,.
\ee
The reciprocal frame $\{e^j\}$ is defined by
\be
e^j\cd e_i=\delta_i^j\,,~~~~\forall i, j=1, \cds, n\,.
\ee
By using the volume element \eqref{vol}, the vectors in the reciprocal frame can be expressed as
\be\label{recif}
e^j=(-1)^{j-1}e_1\w e_2\w\cds\w\check{e}_j\w\cds\w e_n~E_n^{-1}\,,
\ee
where the check on $\check{e}_j$ means this term is missing from the expression, and $E_n^{-1}$ is the inverse of the volume element, defined as $E_nE_n^{-1}=1$. Any vector $a$ can be expressed as linear combination of frame or reciprocal frame vectors,
\be
a=a^ie_i=a_ie^i\,.
\ee
For a grade-r multivector $A_r$, the following equations are satisfied,
\bea
e_ie^i\cd A_r&=&rA_r\,,\label{e1}\\
e_ie^i\w A_r&=&(n-r)A_r\,,\label{e2}\\
e_iA_re^i&=&(-1)^r(n-2r)A_r\,.\label{e3}
\eea

In geometric algebra, the multivector derivative operator is denoted as $\p_X$, here $X$ is a multivector. Suppose an arbitrary function $F(X)$ of multivector argument $X$, the derivative of $F(X)$ with respect to $X$ in the $A$ direction is defined by~\cite{Hestenes1966, Lasenby:1993ya},
\be
\la A\p_XF(X)\ra=\lim_{\tau\to0}\frac{F(X+\tau A)-F(X)}{\tau}\,,
\ee
where $\la~\ra$ means take the scalar part. Given a frame $\{e_i\}$, the multivector derivative $\p_X$ can be expanded as
\be
\p_X=\sum_{i<\cds<j}e^i\w\cds\w e^j\la(e_j\w\cds\w e_i)\p_X\ra\,.
\ee
Particularly, if we choose $X$ as a position vector $x=x^ie_i$, the multivector derivative is simply the nabla in the linear space,
\be
\nabla=e^k\p_{x^k}\,.
\ee
We use the up-dot notation to denote the object that the derivative acting on. For example,
\be
\p_X(AB)=\dot{\p}_X\dot{A}B+\dot{\p}_XA\dot{B}\,.
\ee
By using the notation of vector derivative, equations \eqref{e1}, \eqref{e2} and \eqref{e3} can be expressed in terms of an arbitrary vector $a$,
\bea
\p_aa\cd A_r&=&rA_r\,,\label{a1}\\
\p_aa\w A_r&=&(n-r)A_r\,,\label{a2}\\
\p_aA_ra&=&(-1)^r(n-2r)A_r\,.\label{a3}
\eea

\subsection{Reflection and rotation}
\begin{figure}[H]
\centering
\includegraphics[scale=0.6]{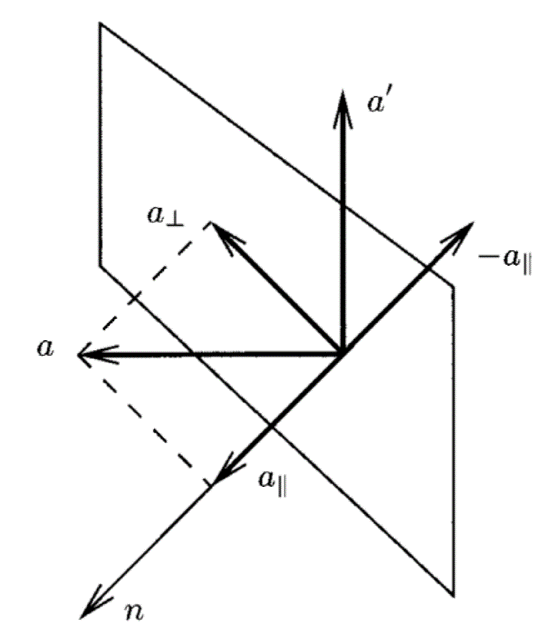}
\caption{Reflecting a vector $a$ in the plane perpendicular to the unit vector $n$.}
\label{figure1}
\end{figure}

In geometric algebra, it is very convenient to do reflection and rotation. Consider an arbitrary vector $a$ and a unit vector $n$ with $n^2=1$. We can decompose the vector $a$ into $a_{||}$ that parallel to $n$ and $a_{\perp}$ that tangent to the plane orthogonal to $n$,
\bea
a=n^2a=n(n\cd a+n\w a)=a_{||}+a_{\perp}\,,
\eea
where
\be
a_{||}=nn\cd a,~~~~a_{\perp}=nn\w a\,.
\ee
As shown in the Figure \ref{figure1}, the result of reflecting $a$ in the plane perpendicular to n is the vector,
\bea
a'=a_{\perp}-a_{||}&=&nn\w a-nn\cd a\nonumber\\
&=&-n\cd an-n\w an\nonumber\\
&=&-nan\,,
\eea
This is a quite neat formula for reflecting vectors along the plane orthogonal to $n$. Under reflection, each vectors in a homogenous multivector $A_r$ transforms according to the above formula, and it is straightforward to show that the reflected $A_r$ takes the form,
\be
A'_r=(-1)^{r}nA_rn\,.
\ee

\begin{figure}[H]
\centering
\includegraphics[scale=0.6]{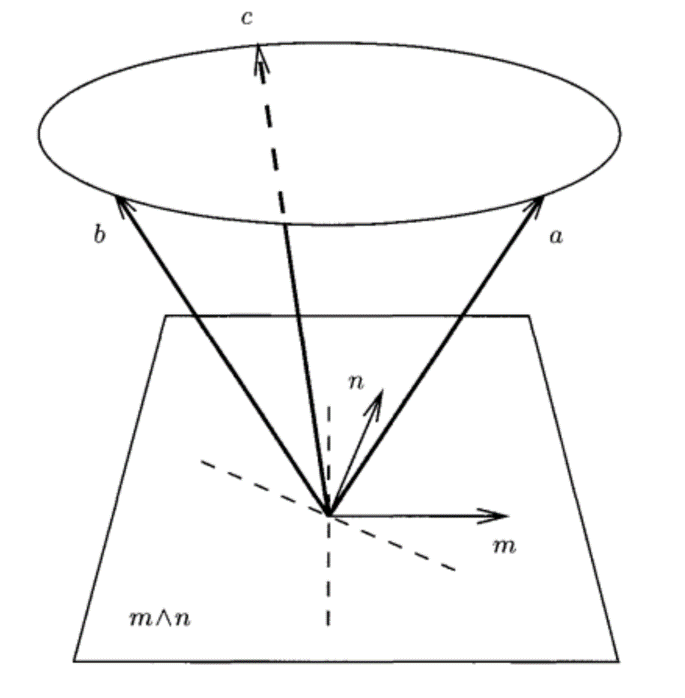}
\caption{Rotation of a vector $a$ in the plane generated by two unit vectors $m$ and $n$. The vector $b$ is the result of reflecting $a$ in the plane perpendicular to $m$, and $c$ is the result of reflecting $b$ in the plane perpendicular to $n$.}
\label{figure2}
\end{figure}

The rotation can be figured out by notice that a rotation in the plane generated by two unit vectors $m$ and $n$ is achieved by successive reflection in the hyperplane perpendicular to $m$ and $n$. See Figure \ref{figure2}. Operably, under rotation, a vector $a$ transforms according to
\be
a\mapsto c=nmamn\,.
\ee
The rotated angle $\theta$ in the $m\w n$ plane is determined by the rotor,
\be
R=nm=e^{-\hat{B}\theta/2}\,,
\ee
where
\be
\cos(\theta/2)=m\cd n,~~~~\hat{B}=\frac{m\w n}{\sin(\theta/2)}\,.
\ee
In geometric algebra, a rotation is generated by a rotor $R$, which is the geometric product of two normalised vectors. Under rotation, any vector $a$ transforms in the following way,
\be
a\mapsto a'=RaR^{\dagger}\,,
\ee
where $\dagger$ means reverse the order in geometric product. Noticing that $R^{\dagger}R=1$, it is straightforward to show that the transformation law of any homogenous multivector $A_r$ under rotation is,
\be\label{rotation}
A_r\mapsto A'_r=RA_rR^{\dagger}\,,
\ee
When the two normalised vector $m$ and $n$ are coincide, the rotor $R$ is the identity and the rotation is trivial.

\subsection{Linear functions}
A linear transformation on the vector space is a map from the original space to itself. In detail, a linear function $\mr{F}$ is a quantity which maps vectors to vectors linearly in the same space. If $a$ is a vector in the space, then $\mr{F}(a)$ is also a vector lies in the same space. The linearity of $\mr{F}$ is ensured by requiring,
\be
\mr{F}(\lambda a+\mu b)=\lambda\mr{F}(a)+\mu\mr{F}(b)\,.
\ee
The transformation of a blade under $\mr{F}$ can be defined by tracing the behaviour of every vector under $\mr{F}$,
\be
\mr{F}(a\w b\w\cds\w c)=\mr{F}(a)\w\mr{F}(b)\w\cds\w\mr{F}(c)\,.
\ee
Given a linear function $\mr{F}$, the adjoint $\bar{\mr{F}}$ is defined as
\be\label{adj}
a\cd\bar{\mr{F}}(b)=\mr{F}(a)\cd b\,,
\ee
for all vectors $a$ and $b$. The adjoint $\bar{\mr{F}}$ is still a linear function. Consider two bivectors $a_1\w a_2$ and $b_1\w b_2$, by using \eqref{useful3}, it is easy to show that,
\bea
(a_1\w a_2)\cd\mr{F}(b_1\w b_2)&=&a_1\cd\mr{F}(b_2)a_2\cd\mr{F}(b_1)-a_1\cd\mr{F}(b_1)a_2\cd\mr{F}(b_2)\nonumber\\
&=&\bar{\mr{F}}(a_1)\cd b_2\bar{\mr{F}}(a_2)\cd b_1-\bar{\mr{F}}(a_1)\cd b_1\bar{\mr{F}}(a_2)\cd b_2\nonumber\\
&=&\bar{\mr{F}}(a_1\w a_2)\cd(b_1\w b_2)\,.
\eea
For arbitrary multivectors $A$ and $B$, the above result can be generalised to
\be
\la A\bar{\mr{F}}(B)\ra=\la\mr{F}(A)B\ra\,.
\ee
And one can also figure out the following useful results for homogenous multivectors,
\bea
A_r\cd\bar{\mr{F}}(B_s)&=&\bar{\mr{F}}(\mr{F}(A_r)\cd B_s),~~~~r\le s\,,\label{F1}\\
\mr{F}(A_r)\cd B_s&=&\mr{F}(A_r\cd\bar{\mr{F}}(B_s)),~~~~r\ge s\,.\label{F2}
\eea

The determinant of a linear function $\mr{det}(\mr{F})$ is defined as
\be\label{det}
\mr{F}(E_n)=\mr{det}(\mr{F})E_n\,,
\ee
where $E_n$ stands for the volume element of the $n$ dimensional vector space. One can show that adjoint function $\bar{\mr{F}}$ has the same determinant of $\mr{F}$ by using the fact that $E_n$ and $E_n^{-1}$ are proportional to each other,
\be
\mr{det}(\bar{\mr{F}})=\la E_n\bar{\mr{F}}(E_n^{-1})\ra=\la \mr{F}(E_n)E_n^{-1}\ra=\mr{det}(\mr{F})\,.
\ee
The inverse of the linear function $\mr{F}$ which is denoted as $\mr{F}^{-1}$ can be figured out by using the determinant,
\be\label{inverseF}
\mr{F}^{-1}(A)=\frac{E_n\bar{\mr{F}}(E_n^{-1}A)}{\mr{det}(\mr{F})}\,.
\ee
where $A$ is an arbitrary multivector.

\section{Gauge Theory Gravity}\label{sec3}
In this section, we aim to model gravitational interactions in terms of gauge fields defined in the geometric algebra for flat spacetime, i.e., spacetime algebra. Let us think a little about our spacetime physics. In our understanding of general relativity, all physical quantities in spacetime correspond to fields, and the gravitational interactions between fields are realised by curving the background spacetime. However, the absolute positions and orientations of the fields in the background spacetime is not measurable. It will drops out of all physical calculations. The things that matter in the physical results are relative relations between fields. When the relative relations are covariantly changed, the physical equations should also change covariantly. In general relativity, this can be realised by writing equations into tensor forms and the tensors are naturally covariant under coordinate changing. To compare the physical fields in different spacetime regions, we use the connection to link and parallelly transport them.

In gauge theory gravity, we keep our background spacetime "flat". The covariance for physical equations are guaranteed by introducing gauge fields. By requiring that the physical system is independent of the position where it is placed and the orientation where it is directed, position gauge fields and rotation gauge fields should be included. Using the language of geometric algebra, the position gauge field is introduced as a linear function $\bar{\mr{h}}(a)$, and it ensures physical equations are covariantly changed under spacetime displacement. The rotational transformation in geometric algebra is realised by sandwich the physical quantities between the rotor and its reverse. To achieve the rotational covariance, the rotors should commute with derivatives. So each derivative operator in equations should be replaced by covariant derivative which contains a bivector valued linear function, i.e. the rotation gauge field $\Omega(a)$.

\subsection{Spacetime algebra}
To be more explicit, let us first briefly introduce the spacetime algebra in four dimensions. The spacetime algebra is generated by four frame vectors $\{\gamma_0, \gamma_1, \gamma_2, \gamma_3\}$ satisfying the following algebraic relations,
\be
\gamma_0^2=1,~~~\gamma_0\cd\gamma_i=0,~~~\gamma_i\cd\gamma_j=-\delta_{ij}\,,
\ee
where $i, j, \cds$ is spacial indices. The vectors in the reciprocal frame can be found by using \eqref{recif},
\be
\gamma^0=\gamma_0,~~~\gamma^i=-\gamma_i\,.
\ee
A spacetime event locates at the position vector $x$,
\be
x=x^{\mu}\gamma_{\mu}=t\gamma_0+x^i\gamma_i\,,
\ee
where $\mu, \nu, \cds$ is spacetime indices. In terms of geometric product, the frame vectors of the spacetime algebra satisfy,
\be\label{spalg}
\gamma_{\mu}\gamma_{\nu}+\gamma_{\nu}\gamma_{\mu}=2\eta_{\mu\nu},~~~~\eta_{\mu\nu}=\mr{diag\{1, -1, -1, -1\}}\,.
\ee
This indicate that the Dirac matrices is a representation of the spacetime algebra. The grade-4 volume element is,
\be
E_4=\gamma_0\gamma_1\gamma_2\gamma_3\,.
\ee
The frame $\{\gamma_{\mu}\}$ define an explicit basis for the spacetime algebra $\mathcal{G}(1, 3)$ as follows,
\begin{table}[htbp]
\centering
\begin{tabular}{ccccccc}
\hline
1, & $\{\gamma_{\mu}\},$ & $\{\gamma_{\mu}\w\gamma_{\nu}\},$ & $\{E_4\gamma_{\mu}\},$ & $E_4.$ \\
\hline
1 scalar, & 4 vectors, & 6 bivectors, & 4 trivectors, & 1 volume element. \\ \hline
\end{tabular}
\end{table}

\subsection{Gauge principle for gravitation}
In this subsection, we figure out the dynamical variables that describe gravitational interactions. The starting point is that all physics relations in spacetime should have the generic form $\Psi(x)=J(x)$, where $\Psi$ and $J$ are spacetime fields representing physical quantities, and $x$ is the spacetime position vector. Due to the universality of physics laws, the establishment of all physics relations should be  completely independent of where we choose $x$ to place. In practise, we can replace $x$ with any new position $x'=f(x)$, and the relation should also be satisfied as $\Psi(x')=J(x')$. On the other hand, the orientational irrelevance of establishing physics relations requires that if we rotate each of $\Psi$ and $J$ by a rotor $R$ and inverse rotate the position vector by the same amount, the equation $R\Psi(R^{\dagger}xR)R^{\dagger}=RJ(R^{\dagger}xR)R^{\dagger}$ has the same physical content as the original equation $\Psi(x)=J(x)$. These two kinds of gauge redundancies enable us to introducing two kinds of dynamical fields, namely, the position gauge field and the rotation gauge field. The physical quantities in the end should be gauge invariant.

\subsubsection{The position gauge fields}
The position gauge field can be introduced by considering the vector derivative $\nabla\phi(x)$ of a scalar field $\phi(x)$. A scalar field $\phi(x)$ is defined as which has spacetime displacement covariance,
\be
\phi'(x)=\phi(x')\,,
\ee
where
\be
x'=f(x)
\ee
is the spacetime displacement and $f(x)$ is an arbitrary differential map between spacetime position vectors. If we act a derivative on the new field $\phi'(x)$ along a vector $a$, we have,
\bea
a\cd\nabla_x\phi'(x)&=&\lim_{\epsilon\to0}\frac{1}{\epsilon}\left(\phi(f(x+\epsilon a))-\phi(f(x))\right)\nonumber\\
&=&\lim_{\epsilon\to0}\frac{1}{\epsilon}\left(\phi(x'+\epsilon\mr{f}(a))-\phi(x')\right)\nonumber\\
&=&\mr{f}(a; x)\cd\nabla_{x'}\phi(x')\,,\label{dd}
\eea
where
\be
\mr{f}(a; x)=a\cd\nabla_xf(x)\,,
\ee
is a linear function of $a$ and also depends on $x$. From the definition of adjoint function \eqref{adj} and
the above equation \eqref{dd}, we find that,
\be
\nabla_x=\bar{\mr{f}}(\nabla_{x'})\,.
\ee
The appearance of $\bar{\mr{f}}$ changes the form of equation contain vector derivative operators. To achieve covariance, we introduce the position gauge field denoted as $\bar{\mr{h}}(a; x)$. It is a linear function of vector $a$ and an arbitrary function of position vector $x$. Under the displacement $x\mapsto x'=f(x)$, this gauge field transforms according to
\be\label{hbdt}
\bar{\mr{h}}(a; x)\mapsto\bar{\mr{h}}'(a; x)=\bar{\mr{h}}(\bar{\mr{f}}^{-1}(a); f(x))\,.
\ee
From the above definition, we can see that the vector derivative operator under the action of gauge field transforms covariantly,
\be
\bar{\mr{h}}(\nabla_x; x)\mapsto\bar{\mr{h}}(\bar{\mr{f}}^{-1}(\nabla_x); f(x))=\bar{\mr{h}}(\nabla_{x'}; x')\,.
\ee
This transformation law ensures that if we define a vector $A(x)$ by
\be
A(x)=\bar{\mr{h}}(\nabla\phi(x))\,,
\ee
then $A(x)$ transforms covariantly as $A(x)\mapsto A'(x)=A(x')$ under arbitrary spacetime displacement.

\subsubsection{The rotation gauge fields}
To introduce the rotation gauge field, let us first check how vector derivative behaves under local spacetime rotation. For the sake of spacetime displacement covariance, quantities like $\nabla J$ should be modified to $\bar{\mr{h}}(\nabla)\mathcal{J}$, where $\mathcal{J}$ is supposed to have displacement and rotation covariance, that is under local spacetime rotation induced by a rotor $R$, $\mathcal{J}$ transforms according to
\be
\mathcal{J}\mapsto R\mathcal{J}R^{\dagger}\,.
\ee
The task now is to make $\bar{\mr{h}}(\nabla)\mathcal{J}$ fully displacement and rotation covariant. To achieve this, we write $\bar{\mr{h}}(\nabla)=\bar{\mr{h}}(\gamma^{\mu})\p_{\mu}$ and check how does the scalar type partial derivative act on local spacetime rotation,
\be
\p_{\mu}(R\mathcal{J}R^{\dagger})=R\p_{\mu}\mathcal{J}R^{\dagger}+2(\p_{\mu}RR^{\dagger})\times(R\mathcal{J}R^{\dagger})\,,
\ee
where we use $\p_{\mu}RR^{\dagger}+R\p_{\mu}R^{\dagger}=0$, and the $\times$ product is defined as $A\times B=\frac{1}{2}(AB-BA)$. To make the derivative covariant, we therefore add a connection term to $\p_{\mu}$ to construct a covariant derivative operator,
\be
\mathcal{D}_{\mu}=\p_{\mu}+\Omega_{\mu}\times\,.
\ee
Here $\Omega_{\mu}$ is a bivector from the rotation gauge field $\Omega(a; x)$, which is a bivector valued linear function of vector $a$ and an arbitrary function of position vector $x$,
\be
\Omega(\gamma_{\mu}; x)=\Omega_{\mu}\,.
\ee
One can show that the $\times$ product of a bivector is grade preserving, so $\mathcal{D}_{\mu}$ is still a scalar type operator. By adding a connection term, under local spacetime rotation, the covariant derivative transforms according to
\be
\mathcal{D}_{\mu}'(R\mathcal{J}R^{\dagger})=R\mathcal{D}_{\mu}\mathcal{J}R^{\dagger}\,,
\ee
given that the connection transforms like
\be\label{Ort}
\Omega(a; x)\mapsto\Omega'(a; x)=R\Omega(a; x)R^{\dagger}-2a\cd\nabla RR^{\dagger}\,.
\ee
Further by requiring that under displacement, $\Omega(a; x)$ transforms according to
\be\label{Odt}
\Omega(a; x)\mapsto\Omega'(a; x)=\Omega(\mr{f}(a); f(x))\,,
\ee
the quantity with the following form,
\be
\mathcal{D}\mathcal{J}=\bar{\mr{h}}(\gamma^{\mu})\mathcal{D}_{\mu}\mathcal{J}\,,
\ee
is fully covariant under displacement and rotation given $\bar{\mr{h}}(a)$ is a vector and transforms like
\be\label{hbrt}
\bar{\mr{h}}(a)\mapsto\bar{\mr{h}}'(a)=R\bar{\mr{h}}(a)R^{\dagger}\,,
\ee
under rotation. Here $\mathcal{D}=\bar{\mr{h}}(\gamma^{\mu})\mathcal{D}_{\mu}$ is the covariant vector derivative, which can also be expressed in terms of a general vector $a$,
\be
\mathcal{D}=\bar{\mr{h}}(\gamma^{\mu})\mathcal{D}_{\mu}=\bar{\mr{h}}(\p_a)\mathcal{D}_a\,,
\ee
where
\be
\mathcal{D}_a=a\cd\nabla+\Omega(a)\times\,.
\ee
The inner product of vector $a$ and $\mathcal{D}$ is a scalar which should have displacement covariance. By using \eqref{a1}, we can write,
\bea
a\cd\mathcal{D}&=&a\cd\bar{\mr{h}}(\p_b)(b\cd\nabla+\Omega(b)\times)\nonumber\\
&=&a\cd\bar{\mr{h}}(\nabla)+\omega(a)\times\,,
\eea
where
\be
\omega(a)=\Omega(h(a))\,,
\ee
denotes the displacement covariant rotation gauge field.

\subsubsection{The covariant spaces}
\begin{figure}[H]
\centering
\includegraphics[scale=0.5]{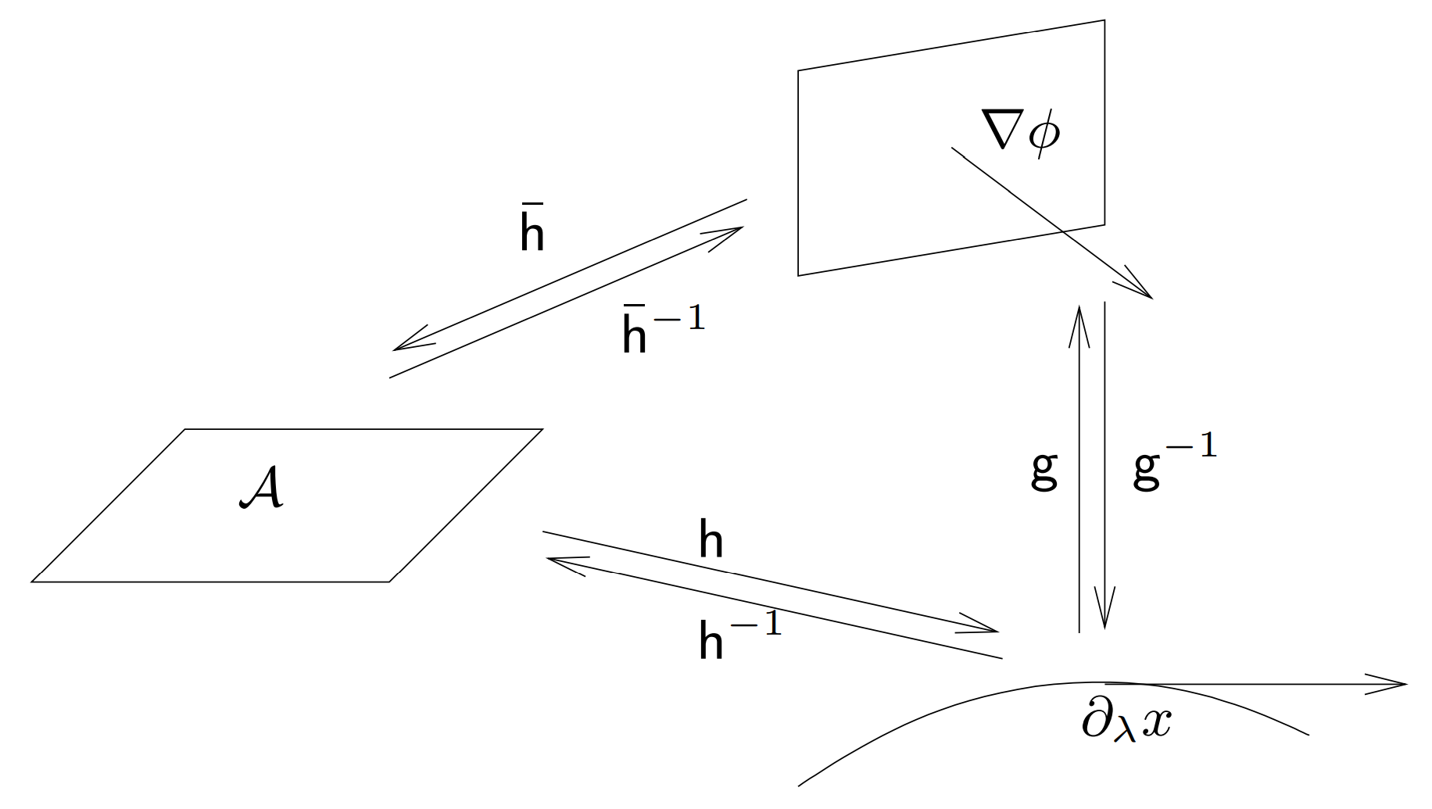}
\caption{The tangent space consisting of tangent vectors $\p_{\lambda}x$, the cotangent space consisting cotangent vectors $\nabla\phi$ and the covariant space consisting of covariant fields $\mathcal{A}$. The metric tensor maps between tangent and cotangent spaces and the gauge field maps between covariant and noncovariant spaces.}
\label{figure3}
\end{figure}

So far, we have introduced two kinds of gauge fields: the vector valued linear function position gauge field $\bar{\mr{h}}(a; x)$ and the bivector valued linear function rotation gauge field $\Omega(a; x)$. This give us $4\times4+4\times6$ degrees of freedom to tackle gravitational interactions in four dimensions. To compare the tensor fields in general relativity, we construct the vectors,
\be\label{gh}
g_{\mu}=\mr{h}^{-1}(\gamma_{\mu}),~~~~g^{\mu}=\bar{\mr{h}}(\gamma^{\mu})\,.
\ee
The metric tensor is defined by
\be\label{g}
g_{\mu\nu}=g_{\mu}\cd g_{\nu}\,.
\ee
The $\bar{\mr{h}}$-field here can be used to construct a vierbein. However, in gauge theory gravity, we should keep in mind that the $\bar{\mr{h}}$-field is used to ensure covariance under displacement. Although we can construct the metric, the background spacetime is still a flat one endowed with the spacetime algebra. In fact, we can now distinct three spaces from flat spacetime, see Figure \ref{figure3}. These are called tangent, cotangent and covariant spaces. The tangent space consists of tangent vectors $a$ expanded by $\gamma_{\mu}$ . The inner products between them are not gauge invariant. The cotangent space consists of cotangent vectors $a^*$ expanded by $\gamma^{\mu}$. The vectors in tangent and cotangent spaces can be interchanged via the metric tensor, which maps one space to the other, and this map can be write in a frame free form,
\be
a^*=\bar{\mr{h}}^{-1}\mr{h}^{-1}(a)=g(a)\,.
\ee
The tangent and cotangent spaces, as well as the metric map between them, are the traditional elements of general relativity. The covariant space consists of covariant fields who transform covariantly under displacement and rotation. This is unique to the gauge theory formulation. The physical quantities including gravitational field are defined as gauge invariant quantities in the covariant space.

The gauge formulation of gravity with geometric algebra have many advantages. The gravitational field equations written in gauge covariant form lies within the covariant spaces. Compare to the tensor manipulation in general relativity, the application of spacetime algebra in gauge theory gravity simplifies the calculation. And conceptually,  the physical relevant objects become much clear due to the gauge invariance.

\subsection{The gravitational field equations}
In gauge theory gravity, the dynamical gravitational fields are introduced through gauge covariance. Similar to the electromagnetism, the field equations can be constructed from an action made by the field strength. By the same token, the field strength for gauge gravity is defined by the commutator of the covariant derivative operator,
\be
[\mathcal{D}_a,~\mathcal{D}_b]M=R(a\w b)M\,,
\ee
where $M$ is any multivector field. The gauge gravity field strength $R(B)$ defined in the above formula can be expressed as~\cite{Doran2003, Lasenby:2019gmi}
\be
R(a\w b)=a\cd\nabla\Omega(b)-b\cd\nabla\Omega(a)+\Omega(a)\times\Omega(b)\,.
\ee
Under displacement, $R(B)$ transforms as
\be
R(B; x)\mapsto R'(B; x)=R(\mr{f}(B); x')\,.
\ee
And the rotational transformation law of the field strength is,
\be
R(B)\mapsto R'(B)=RR(B)R^{\dagger}\,.
\ee
From the above transformations, a covariant field strength therefore can be constructed,
\be\label{cR1}
\mathcal{R}(B; x)=R(\mr{h}(B); x)\,,
\ee
which has the following transformation laws,
\bea
\mr{Displacement:}~~~~&&\mathcal{R}'(B; x)=\mathcal{R}(B; x')\,,\\
\mr{Rotation:}~~~~&&\mathcal{R}'(B)=R\mathcal{R}(R^{\dagger}BR)R^{\dagger}\,.
\eea
Here $\mathcal{R}(B)$ is the gauge theory analogue of the Riemann tensor. From the field strength $\mathcal{R}(B)$, the analogue of Ricci tensor and scalar, and Einstein tensor can be formulated with vector derivative inner product,
\bea
\mr{Ricci~Tansor:}~~~~&&\mathcal{R}(b)=\p_a\cd\mathcal{R}(a\w b)\,,\label{Riccit}\\
\mr{Ricci~Scalar:}~~~~&&\mathcal{R}=\p_b\cd\mathcal{R}(b)\,,\label{Riccis}\\
\mr{Einstein~Tensor:}~~~~&&\mathcal{G}(a)=\mathcal{R}(a)-\frac{1}{2}a\mathcal{R}\,.\label{Einsteint}
\eea
The Ricci scalar is a good candidate for a Lagrangian density of the gravitational gauge fields, since it is displacement covariant and rotation invariant. To construct the action, we also need a gauge invariant scalar measure. In four spacetime dimensions, the volume element is,
\be
E_4=\gamma_0\gamma_1\gamma_2\gamma_3\,.
\ee
Under displacement, these vectors transforms as
\be
\gamma'_{\mu}=\frac{\p f(x)}{\p x^{\mu}}=\mr{f}(\gamma_{\mu})\,.
\ee
Note that the position gauge field has the transformation law \eqref{hbdt} and the definition of the determinant \eqref{det}, the gauge invariant scalar measure can be written as
\be
-\mr{h}^{-1}(\gamma_0)\w\cds\w\mr{h}^{-1}(\gamma_3)E_4dx^0\cds dx^3=\mr{det}(\mr{h}^{-1})d^4x\,.
\ee
The gauge invariant gravitational action can be written down,
\be
S=\int d^4x\mr{det}(\mr{h^{-1}})\left(\frac{1}{2}\mathcal{R}+\Lambda-\kappa\mathcal{L}_m\right)\,,
\ee
where $\mathcal{L}_m$ is the matter Lagrangian density and $\kappa=8\pi G$. $\Lambda$ is an arbitrary constant which can be viewed as a cosmological constant. There are two dynamical variables $\bar{\mr{h}}(a; x)$ and $\Omega(a; x)$ describing the gravitational fields. In terms of these fields, the Ricci scalar takes the form,
\be
\mathcal{R}=\bar{\mr{h}}(\p_b\w\p_a)\cd(a\cd\nabla\Omega(b)-b\cd\nabla\Omega(a)+\Omega(a)\times\Omega(b))\,.
\ee

The Euler-Lagrange equation for the $\bar{\mr{h}}-$field can be derived from the least action principle,
\be\label{ELhb}
\p_{\bar{\mr{h}}(a)}\left(\mr{det}(\mr{h}^{-1})\left(\frac{\mathcal{R}}{2}+\Lambda-\kappa\mathcal{L}_m\right)\right)=0\,.
\ee
Here we assume that $\bar{h}(a)$ and $\Omega(a)$ appears undifferentiated in the matter part. The details of doing partial derivative with respect to $\bar{\mr{h}}(a)$ and $\Omega(a)$ can be found in Appendix A. The result we can get is the following,
\be
\p_{\bar{\mr{h}}(a)}\left(\frac{1}{2}\mr{det}(\mr{h}^{-1})\mathcal{R}+\mr{det}(\mr{h}^{-1})\Lambda\right)=\frac{1}{\mr{det}(\mr{h})}\left(\p_b\cd\mathcal{R}(b\w\mr{h}^{-1}(a))-\frac{\mathcal{R}+2\Lambda}{2}\mr{h}^{-1}(a)\right)\,.
\ee
We define the covariant stress energy tensor $\mathcal{T}$(a) as following,
\be
\mathcal{T}(\mr{h}^{-1}(a))=\mr{det}(\mr{h})\p_{\bar{\mr{h}}(a)}(\mathcal{L}_m\mr{det}(\mr{h}^{-1}))\,.
\ee
Thus from the Euler-Lagrange equation \eqref{ELhb}, we obtain the equation,
\be\label{Einsteineq}
\mathcal{G}(a)-\Lambda a=\kappa\mathcal{T}(a)\,,
\ee
where $\mathcal{G}(a)$ is the Einstein tensor defined in \eqref{Einsteint}. This is the gauge theory analogue of Einstein's equation.

The Euler-Lagrange equation for the $\Omega-$field is,
\be\label{ELO}
\p_{\Omega(a)}\mathcal{R}-\mr{det}(\mr{h})\p_b\cd\nabla(\p_{\Omega(a), b}\mathcal{R}\mr{det}(\mr{h}^{-1}))=2\kappa\p_{\Omega(a)}\mathcal{L}_m\,,
\ee
here we use the comma symbol to denote the common derivative, i.e., $\Omega(a), b=b\cd\nabla\Omega(a)$. Using the results in Appendix A, we can find,
\be
\p_{\Omega(a)}\mathcal{R}-\mr{det}(\mr{h})\p_b\cd\nabla(\p_{\Omega(a), b}\mathcal{R}\mr{det}(\mr{h}^{-1}))=\mathcal{D}\w\bar{\mr{h}}(a)+\mr{det}(\mr{h})\mathcal{D}_{\p_b}(\bar{\mr{h}}(b)\mr{det}(\mr{h}^{-1}))\w\bar{\mr{h}}(a)\,.
\ee
We define the spin of the matter as following,
\be
S(a)=\p_{\Omega(a)}\mathcal{L}_m\,.
\ee
So the $\Omega-$field equation takes the form,
\be\label{Oeq}
\mathcal{D}\w\bar{\mr{h}}(a)+\mr{det}(\mr{h})\mathcal{D}_{\p_b}(\bar{\mr{h}}(b)\mr{det}(\mr{h}^{-1}))\w\bar{\mr{h}}(a)=\kappa S(a)\,.
\ee
We can further simplify the above equation by doing inner product with $\mr{h}^{-1}(\p_a)$ on both sides. The first term on the right hand side of \eqref{Oeq} will leads to
\bea
\mr{h}^{-1}(\p_a)\cd(\mathcal{D}\w\bar{\mr{h}}(a))&=&\mr{h}^{-1}(\p_a)\cd(\bar{\mr{h}}(\p_b)\w b\cd\nabla\bar{\mr{h}}(a))+\mr{h}^{-1}(\p_a)\cd(\bar{\mr{h}}(\p_b)\w(\Omega(b)\times\bar{\mr{h}}(a)))\nonumber\\
&=&\mr{det}(\mr{h})\p_b\cd\nabla(\bar{\mr{h}}(b)\mr{det}(\mr{h}^{-1}))+\Omega(\p_b)\times\bar{\mr{h}}(b)\nonumber\\
&=&\mr{det}(\mr{h})\mathcal{D}_{\p_b}(\bar{\mr{h}}(b)\mr{det}(\mr{h}^{-1}))\,,
\eea
while from the second term, we can get,
\be
\mr{det}(\mr{h})\mr{h}^{-1}(\p_a)\cd(\mathcal{D}_{\p_b}(\bar{\mr{h}}(b)\mr{det}(\mr{h}^{-1}))\w\bar{\mr{h}}(a))=-3\mr{det}(\mr{h})\mathcal{D}_{\p_b}(\bar{\mr{h}}(b)\mr{det}(\mr{h}^{-1}))\,.
\ee
In all, we have the following relation,
\be\label{rela}
\mr{det}(\mr{h})\mathcal{D}_{\p_b}(\bar{\mr{h}}(b)\mr{det}(\mr{h}^{-1}))=-\frac{1}{2}\kappa\mr{h}^{-1}(\p_b)\cd S(b)\,.
\ee
Substituting \eqref{rela} into \eqref{Oeq}, we can get,
\be\label{torsioneq}
\mathcal{D}\w\bar{\mr{h}}(a)=\kappa\left(S(a)+\frac{1}{2}(\mr{h}^{-1}(\p_b)\cd S(b))\w\bar{\mr{h}}(a)\right)\,.
\ee
This is the second of our gravitational equation. $\mathcal{D}\w\bar{\mr{h}}(a)$ is identified as the gravitational torsion, and \eqref{torsioneq} is the torsion equation which relate the gravitational torsion to the matter spin.

In the case of matter with vanishing spin, the torsion equation \eqref{torsioneq} can be expressed as the torsion free condition,
\be\label{teqs=0}
\bar{\mr{h}}(\dot{\nabla})\w\dot{\bar{\mr{h}}}(a)=-\p_b\w(\omega(b)\cd\bar{\mr{h}}(a))\,,
\ee
where the updot indicate the scope of derivative. This equation can be used to find out the relation between $\bar{\mr{h}}(a)$ and $\omega(a)$ in the spinless case. For this purpose, we define a bivector valued linear function,
\be\label{H}
\mr{H}(a)=\bar{\mr{h}}(\dot{\nabla}\w\dot{\bar{\mr{h}}}^{-1}(a))=-\bar{\mr{h}}(\dot{\nabla})\w\dot{\bar{\mr{h}}}(\bar{\mr{h}}^{-1}(a))\,.
\ee
In terms of $\mr{H}(a)$, \eqref{teqs=0} can be written as
\be
\p_b\w(\omega(b)\cd a)=\mr{H}(a)\,.
\ee
By successively doing outer product with $\p_a$ and inner product with $a$ on both sides of the above equation, we can get,
\be\label{wa}
\omega(a)=-\mr{H}(a)+\frac{1}{2}a\cd(\p_b\w\mr{H}(b))\,.
\ee
Note that $\bar{\mr{h}}(\nabla)$ and $\omega(a)$ are both with displacement covariance, these are the quantities with which we would like to express the physical fields. In the torsion free case, the covariant field strength \eqref{cR1} can be explicitly written in terms of $\bar{\mr{h}}(\nabla)$ and $\omega(a)$. We therefore define the operator,
\be
L_a=a\cd\bar{\mr{h}}(\nabla)\,.
\ee
By noticing that the torsion free equation \eqref{teqs=0} also implies,
\be
\la(b\w a)(\bar{\mr{h}}(\dot{\nabla})\w\dot{\bar{\mr{h}}}(c))\ra=-\la(b\w a)(\p_d\w(\omega(d))\cd\bar{\mr{h}}(c))\ra\,,
\ee
thus the intrinsic content of \eqref{teqs=0} can be gathered in the following commutator relation,
\be
[L_a, L_b]=(L_ab-L_ba)\cd\bar{\mr{h}}(\nabla)=(\dot{L}_a\dot{\mr{h}}(b)-\dot{L}_b\dot{\mr{h}}(a))\cd\nabla=L_c\,,
\ee
where
\be
c=a\cd\mathcal{D}b-b\cd\mathcal{D}a\,.
\ee
The covariant field strength \eqref{cR1} now can be expressed as
\bea\label{cR2}
\mathcal{R}(a\w b)&=&\dot{L}_a\dot{\Omega}(\mr{h}(b))-\dot{L}_b\dot{\Omega}(\mr{h}(a))+\omega(a)\times\omega(b)\nonumber\\
&=&L_a\omega(b)-L_b\omega(a)+\omega(a)\times\omega(b)-\omega(c)\,.
\eea

Invoking equation \eqref{torsioneq}, when the spin of matter vanishes, we can derive that for any multivector $A$,
\be
\mathcal{D}\w A=\mathcal{D}\w\bar{\mr{h}}(\bar{\mr{h}}^{-1}(A))=0\,.
\ee
And hence,
\bea
\mathcal{D}\w\mathcal{D}\w A&=&\bar{\mr{h}}(\p_a)\w\bar{\mr{h}}(\p_b)\w(\mathcal{D}_a\mathcal{D}_bA)\nonumber\\
&=&\frac{1}{2}\bar{\mr{h}}(\p_a)\w\bar{\mr{h}}(\p_b)\w(R(a\w b)A)\nonumber\\
&=&\frac{1}{2}\p_a\w\p_b\w(\mathcal{R}(a\w b)A)\nonumber\\
&=&0
\eea
Since $A$ is an arbitrary multivector, we have,
\be\label{Rid1}
\p_a\w\p_b\w\mathcal{R}(a\w b)=0\,.
\ee
By doing a inner product with any vector $c$ on the above equation, we can further get,
\be\label{Rid2}
\p_a\w\mathcal{R}(a\w b)=0\,.
\ee
\eqref{Rid1} and \eqref{Rid2} are the mathematical identities satisfied by the covariant field strength. In the gauge theory gravity, the covariant Weyl tensor $\mathcal{W}(a\w b)$ is a bivector defined as
\be\label{Weyl}
\mathcal{W}(a\w b)=\mathcal{R}(a\w b)-\frac{1}{2}(\mathcal{R}(a)\w b+a\w\mathcal{R}(b))+\frac{1}{6}\mathcal{R}a\w b
\ee
in four spacetime dimensions. With the identities satisfied by $\mathcal{R}(a\w b)$, one can show that the covariant Weyl tensor satisfies,
\bea
\p_a\cd\mathcal{W}(a\w b)&=&0\,,\\
\p_a\w\mathcal{W}(a\w b)&=&0\,.
\eea
The expression \eqref{Weyl} states that we can decompose the covariant field strength into purely gravitational Weyl part and the rest terms are solely determined by the stress energy tensor from the matter.

\section{Exact Gravitational Waves with $\Lambda<0$}\label{sec4}
In the gauge theory gravity, the gravitational fields are represented by two gauge covariant linear function $\bar{\mr{h}}(a)$ and $\omega(a)$, and the dynamics of them are governed by field equations \eqref{Einsteineq} and \eqref{torsioneq}. In this section, we present the exact gravitational wave solutions of these equations when $\Lambda$ is a negative constant.

In four spacetime dimensions, the frame vectors $\{\gamma_{\mu}\}$ satisfy the spacetime algebra \eqref{spalg}. The position vector $x$ has the corresponding coordinates $\{t, r, y, z\}$ in this frame,
\be
x=t\gamma_0+r\gamma_1+y\gamma_2+z\gamma_3\,.
\ee
We are looking for a gravitational wave propagating in the $\gamma_3$ direction. We define $\gamma_{\pm}=\frac{1}{\sqrt{2}}(\gamma_0\pm\gamma_3)$, and hence they satisfy,
\be\label{pm}
\gamma_+\cd\gamma_-=\gamma_-\cd\gamma_+=1,~~~~\gamma_+\cd\gamma_+=\gamma_-\cd\gamma_-=0\,.
\ee
The ansatz we choose for the position gauge field $\bar{\mr{h}}(a)$ is as following,
\be\label{az}
\bar{\mr{h}}(a)=\frac{r}{\ell}a-\frac{r}{2\ell}H(t, r, y, z)(\gamma_+\cd a)\gamma_+\,,
\ee
where $\ell$ is a positive constant and $H(t, r, y, z)$ is a undetermined function of spacetime coordinates. By acting $\bar{\mr{h}}^{-1}$ on both sides of \eqref{az} and using \eqref{pm}, it can be shown that,
\be
\bar{\mr{h}}^{-1}(a)=\frac{\ell}{r}a+\frac{\ell}{2r}H(t, r, y, z)(\gamma_+\cd a)\gamma_+\,.
\ee
Similarly, one can also show that,
\bea
\mr{h}(a)&=&\frac{r}{\ell}a-\frac{r}{2\ell}H(t, r, y, z)(\gamma_+\cd a)\gamma_+\,,\\
\mr{h}^{-1}(a)&=&\frac{\ell}{r}a+\frac{\ell}{2r}H(t, r, y, z)(\gamma_+\cd a)\gamma_+\,.
\eea
We assume the gravitational wave is travelling in the vacuum spacetime without matter, so we can use the torsion free condition \eqref{teqs=0} to express $\omega(a)$ in terms of $\bar{\mr{h}}(a)$ through \eqref{wa}. Notice that
\bea
\dot{\nabla}\w\dot{\bar{\mr{h}}}^{-1}(a)&=&\nabla\left(\frac{\ell}{r}\right)\w a+\frac{\ell}{2}(\gamma_+\cd a)\nabla\left(\frac{H}{r}\right)\w\gamma_+\nonumber\\
&=&\frac{\ell}{r^2}\gamma_1\w a+\frac{\ell}{2}(\gamma_+\cd a)\nabla\left(\frac{H}{r}\right)\w\gamma_+\,,
\eea
and act on the linear function $\bar{\mr{h}}$ by using \eqref{az}, we can find,
\bea
\mr{H}(a)&=&\bar{\mr{h}}(\dot{\nabla}\w\dot{\bar{\mr{h}}}^{-1}(a))=\frac{\ell}{r^2}\bar{\mr{h}}(\gamma_1)\w\bar{\mr{h}}(a)+\frac{\ell}{2}(\gamma_+\cd a)\bar{\mr{h}}\left(\nabla\left(\frac{H}{r}\right)\right)\w\bar{\mr{h}}(\gamma_+)\nonumber\\
&=&\frac{1}{\ell}\gamma_1\w a+\frac{r}{2\ell}(\gamma_+\cd a)\nabla H\w\gamma_+\,.
\eea
It is easy to show that $\p_b\w\mr{H}(b)=0$, since $\p_b\w b=0$, $\p_b(\gamma_+\cd b)=\gamma_+$ and $\gamma_+\w\gamma_+=0$. Then from \eqref{wa}, the covariant rotation gauge field takes the form,
\be\label{waH}
\omega(a)=-\frac{1}{\ell}\gamma_1\w a-\frac{r}{2\ell}(\gamma_+\cd a)\nabla H\w\gamma_+\,.
\ee
The function $H(t, r, y, z)$ characterise the magnitude and the the way of propagating of the gravitational wave. We set $H(t, r, y, z)=H(t-z, r, y)$ to represent a monochromatic gravitational wave propagating in the $\gamma_3$ direction. Under this assumption, each terms in the covariant field strength \eqref{cR2} can be shown,
\bea
a\cd\bar{\mr{h}}(\nabla)\omega(b)&=&-\frac{r}{2\ell^2}(\gamma_+\cd b)a\cd\nabla(r\nabla H\w\gamma_+)\,,\label{Rab1}\\
\omega(a)\times\omega(b)&=&\frac{1}{\ell}a\w(\gamma_1\cd\omega(b))-\frac{1}{\ell}\gamma_1\w(a\cd\omega(b))-\frac{r}{2\ell^2}(\gamma_+\cd a)(\gamma_1\w b)\times(\nabla H\w\gamma_+)\,,\nonumber\\\label{Rab2}
\\
\omega(c)&=&\omega(\omega(a)\cd b-\omega(b)\cd a)\nonumber\\
&=&\frac{1}{\ell}(\gamma_1\cd b)\omega(a)-\frac{r}{2\ell}(\gamma_+\cd b)(\nabla H\cd a)\omega(\gamma_+)-(a\leftrightarrow b)\,.\label{Rab3}
\eea
The Ricci tensor \eqref{Riccit} is a grade-1 vector obtained by doing inner product with $\p_a$ and $\mathcal{R}(a\w b)$. The ingredients are,
\bea
\p_a\cd(a\cd\bar{\mr{h}}(\nabla)\omega(b))&=&-\frac{r}{2\ell^2}(\gamma_+\cd b)\nabla\cd(r\nabla H\w\gamma_+)\,,\\
\p_a\cd(b\cd\bar{\mr{h}}(\nabla)\omega(a))&=&0\,,\\
\p_a\cd(\omega(a)\times\omega(b))&=&\frac{2}{\ell}\gamma_1\cd\omega(b)-\frac{r}{2\ell^2}\gamma_+\cd((\gamma_1\w b)\times(\nabla H\w\gamma_+))\nonumber\\
&=&\frac{2}{\ell^2}((\gamma_1\cd b)\gamma_1+b)-\frac{r}{2\ell^2}(\gamma_+\cd b)(\nabla H\cd\gamma_1)\gamma_+\,,\\
\p_a\cd\omega(c)&=&\frac{2}{\ell^2}\left((\gamma_1\cd b)\gamma_1-\frac{b}{2}\right)+\frac{r}{\ell^2}(\gamma_+\cd b)(\nabla H\cd\gamma_1)\gamma_+\,.
\eea
Combine them together, we find,
\be\label{Rb}
\mathcal{R}(b)=\p_a\cd\mathcal{R}(a\w b)=\frac{3}{\ell^2}b-\frac{r^2}{2\ell^2}(\gamma_+\cd b)\left(\nabla^2H+\frac{2}{r}\gamma_1\cd\nabla H\right)\gamma_+\,.
\ee
The Ricci scalar defined in \eqref{Riccis} can be found as a constant,
\be\label{R}
\mathcal{R}=\p_b\cd\mathcal{R}(b)=12/\ell^2\,,
\ee
and the Einstein tensor \eqref{Einsteint} takes the form,
\be
\mathcal{G}(b)=-\frac{3}{\ell^2}b-\frac{r^2}{2\ell^2}(\gamma_+\cd b)\left(\nabla^2H+\frac{2}{r}\gamma_1\cd\nabla H\right)\gamma_+\,.
\ee
Now given that $\Lambda=-3/\ell^2$ is a negative cosmological constant and
\be\label{Heq}
\nabla^2H+\frac{2}{r}\gamma_1\cd\nabla H=0\,,
\ee
the Einstein like field equation of $\bar{\mr{h}}(a)$ \eqref{Einsteineq} is satisfied. In terms of spacetime coordinates, the equation of $H(t-z, r, y)$ can be expressed as
\be\label{Hsol}
\left(\frac{\p^2}{\p r^2}+\frac{\p^2}{\p y^2}\right)H(t-z, r, y)-\frac{2}{r}\frac{\p}{\p r}H(t-z, r, y)=0\,.
\ee
Such kind of equation also appears in the Siklos spacetime in general relativity~\cite{Siklos1985, Podolsky:1997ni, Podolsky:1997ik}. The explicit solutions to this equation can be written as
\be\label{Hsolex}
H=r^2\frac{\p}{\p r}\left(\frac{\xi+\bar{\xi}}{r}\right)\,,
\ee
where $\xi=\xi(t-z, r+iy)$ is an arbitrary analytic function on the $r+iy$ plane.

Since we are looking for a gravitational wave which is a purely radiation solution without matter, we should check that the Weyl tensor defined in \eqref{Weyl} of the gravitational wave is type-N in the Petrov classification, that is~\cite{Lasenby:1998yq},
\be\label{tN}
\mathcal{W}^2(a\w b)=0\,.
\ee
Combine all the terms in \eqref{Rab1}, \eqref{Rab2}, \eqref{Rab3}, \eqref{Rb} and \eqref{R} together and after some manipulations, we can write the Weyl tensor in a compact form expressed in the geometric product,
\bea
\mathcal{W}(a\w b)&=&\frac{r^2}{4\ell^2}((\gamma_+\cd a)(\gamma_+\w b)-(\gamma_+\cd b)(\gamma_+\w a))\nabla^2H\nonumber\\
&+&\frac{r^2}{2\ell^2}((\gamma_+\cd a)(b\cd\nabla)-(\gamma_+\cd b)(a\cd\nabla))(\nabla H\w\gamma_+)\nonumber\\
&=&-\frac{r^2}{8\ell^2}\gamma_+\nabla((a\w b)\nabla H)\gamma_+\,.\label{Wab}
\eea
Since $\gamma_+\gamma_+=0$, the covariant Weyl tensor thus satisfies \eqref{tN}, so the gravitational wave is indeed a type-N solution.

The polarization of the gravitational wave specified by the solution \eqref{Hsolex} can be figured out by investigating the geodesic deviation of free falling particles. Suppose a congruence of geodesic with covariant velocity vector field $v$ and separation vector field $n$. These two vector fields should satisfy,
\be
v\cd\mathcal{D}v=0,~~~~~[v,~n]=0\,.
\ee
The geodesic deviation equation, which gives the relative acceleration vector $a$ among free falling particles, can be written in terms of the covariant Weyl tensor and the cosmological constant,
\bea
a&=&v\cd\mathcal{D}(v\cd\mathcal{D}n)\nonumber\\
&=&\mathcal{R}(v\w n)\cd v\nonumber\\
&=&\left(\mathcal{W}(v\w n)+\frac{1}{2}(\mathcal{R}(v)\w n+v\w\mathcal{R}(n))-\frac{1}{6}v\w n\mathcal{R}\right)\cd v\nonumber\\
&=&\mathcal{W}(v\w n)\cd v+\frac{\Lambda}{3}n\,.\label{acc}
\eea
Given the covariant Weyl tensor \eqref{Wab} as well as an explicit solution of $H(t-z, r, y)$ in \eqref{Hsolex}, one can see the deviation among free falling particles which reflects the polarization of the gravitational wave. If we choose the particle velocity along $\gamma_0$ and the separation vector orthogonal to the gravitational wave propagating direction, that is we set the free falling particles initially at rest in the $(r-y)$-plane, and the relative acceleration \eqref{acc} will record the polarization of the gravitational wave propagating in the $\gamma_3$ direction. We can list several examples of explicit solutions and their corresponding covariant Weyl tensors,
\bea
H_1=h_1(t-z)(r^2+y^2)\,,&&\mathcal{W}_1=0\,,\\
H_2=h_2(t-z)r^3\,,&&\mathcal{W}_2=-h_2(t-z)\frac{3r^3}{8\ell^2}\mathcal{W}^+\,,\label{h2}\\
H_3=h_3(t-z)\left(\frac{\cos\theta}{3}r^3+\frac{\sin\theta}{6}y^3+\frac{\sin\theta}{2}r^2y\right)\,,&&\mathcal{W}_3=-h_3(t-z)\frac{r^3}{8\ell^2}\left(\cos\theta\mathcal{W}^++\sin\theta\mathcal{W}^{\times}\right)\,,\label{h3}
\eea
where $\theta$ is a function of $t-z$, and
\bea
\mathcal{W}^+&=&\gamma_+(\gamma_1(v\w n)\gamma_1-\gamma_2(v\w n)\gamma_2)\gamma_+\,,\\
\mathcal{W}^{\times}&=&\gamma_+(\gamma_1(v\w n)\gamma_2+\gamma_2(v\w n)\gamma_1)\gamma_+\,,
\eea
are the two polarization modes of the gravitational wave with $\mathcal{W}^+$ and $\mathcal{W}^{\times}$ indicating particles acceleration along and perpendicular to the separation direction in the polarization plane respectively. That is the solution $H_1$ has no polarization mode, the solution $H_2$ has one $+$ mode, and the solution $H_3$ has circular periodic modes between $+$ and $\times$ polarization. The negative cosmological constant in \eqref{acc} will always give a attractive effect among particles moving freely.

\section{Particle Motion and Velocity Memory Effect}\label{sec5}
Let us now consider a test massive particle moving in the presence of the gravitational gauge field $\bar{\mr{h}}(a)$ in the form \eqref{az} with solution \eqref{Hsolex}. Much like a charged particle moving on the background of electromagnetic field. The equation of motion of the particle can be derived from a gauge invariant action,
\be
S=m\int d\lambda\sqrt{\mr{h}^{-1}(\dot{x}(\lambda))\cd\mr{h}^{-1}(\dot{x}(\lambda))}=m\int d\lambda\sqrt{g_{\mu\nu}\dot{x}^{\mu}(\lambda)\dot{x}^{\nu}(\lambda)}\,,
\ee
where $m$ is the mass of the particle, $\lambda$ is the affine parameter along the particle's worldline and the updot label the derivative with respect to $\lambda$. We use $x(\lambda)$ to denote the position vector of the particle and $x^{\mu}(\lambda)$ are its spacetime coordinates. We have invoked \eqref{gh} and \eqref{g} to expressed the action in a form with metric. By doing variation with respect to $x(\lambda)$, the least action principle can give us the equation of motion of the particle~\cite{Lasenby:1998yq},
\be\label{geod}
v\cd\mathcal{D}v=\dot{v}+\omega(v)\cd v=0\,,
\ee
where $v=\mr{h}^{-1}(\dot{x})$ is the spacetime covariant velocity of the particle, and the timelike requirement of the massive particle is ensured by $v\cd v=1$.

On the background of the gravitational wave, the particle's covariant velocity takes the form,
\be
v=\frac{\ell}{r}\dot{x}+\frac{\ell}{2r}H(t-z, r, y)(\gamma_+\cd\dot{x})\gamma_+\,.
\ee
Substituting the above velocity into the equation of motion \eqref{geod} and using \eqref{waH}, we can find the following equation,
\be
\ddot{x}=\frac{r}{\ell^2}\gamma_1+2\frac{\dot{r}}{r}\dot{x}+\frac{1}{2}(\gamma_+\cd\dot{x})^2\nabla H-(\gamma_+\cd\dot{x})(\nabla H\cd\dot{x})\gamma_+\,.
\ee
In components, the equations of motion of the particle are,
\bea
\ddot{t}&=&2\frac{\dot{r}}{r}\dot{t}+\frac{1}{4}(\dot{t}-\dot{z})^2\p_tH-\frac{1}{2}(\dot{t}-\dot{z})(\nabla H\cd\dot{x})\,,\\
\ddot{z}&=&2\frac{\dot{r}}{r}\dot{z}-\frac{1}{4}(\dot{t}-\dot{z})^2\p_zH-\frac{1}{2}(\dot{t}-\dot{z})(\nabla H\cd\dot{x})\,,\\
\ddot{r}&=&\frac{r}{\ell^2}+2\frac{\dot{r}^2}{r}-\frac{1}{4}(\dot{t}-\dot{z})^2\p_rH\,,\\
\ddot{y}&=&2\frac{\dot{r}}{r}\dot{y}-\frac{1}{4}(\dot{t}-\dot{z})^2\p_yH\,.
\eea
Here we make a set of new coordinates $\{u, v, r, y\}$ with $u=t-z$ and $v=t+z$. In terms of this new coordinates, we have $H=H(u, r, y)$ and the above equations can be reexpressed as,
\bea
\ddot{u}&=&2\frac{\dot{r}}{r}\dot{u}\,,\label{equ}\\
\ddot{v}&=&2\frac{\dot{r}}{r}\dot{v}+\frac{1}{2}\dot{u}^2\p_uH-\frac{1}{2}\dot{u}(\nabla H\cd\dot{x})\,,\label{eqv}\\
\ddot{r}&=&\frac{r}{\ell^2}+2\frac{\dot{r}^2}{r}-\frac{1}{4}\dot{u}^2\p_rH\,,\label{eqr}\\
\ddot{y}&=&2\frac{\dot{r}}{r}\dot{y}-\frac{1}{4}\dot{u}^2\p_yH\,,\label{eqy}
\eea
From \eqref{equ}, we can derive,
\be\label{ur}
\dot{u}=kr^2\,,
\ee
where $k$ is a constant of motion which is related to the light-cone energy of the particle. If we reparameterize the particle worldline by using coordinate $u$ instead of $\lambda$, and invoking relation \eqref{ur}, we can write the $r$ and $y$'s equations into the following form,
\bea
\frac{d^2r}{du^2}&=&\frac{1}{k^2\ell^2r^3}-\frac{1}{4}\p_rH(u, r, y)\,,\label{rdu}\\
\frac{d^2y}{du^2}&=&-\frac{1}{4}\p_yH(u, r, y)\,.\label{ydu}
\eea
For generic $H(u, r, y)$, $r$ and $y$ components are determined by the above two equations.

We consider the velocity memory effect when a gravitational impulsive wave had passed over the particle, the change in velocity of the particle hence will record the wave information carried by the pulse. We first take the solution $H_2(u, r)=h_2(u)r^3$ as in \eqref{h2}, and for gravitational impulsive wave, the $u$ dependence can be written as
\be\label{pulse}
h_2(u)=
\begin{cases}
Ae^{-bu}, \quad \ \  &u>0\\
Ae^{bu},\quad \ \  &u<0
\end{cases}\,,
\ee
where $A$ is the pulse amplitude and $b$ is a positive constant. The center of this impulsive wave is at $u=0$ and it is propagating in the $\gamma_3$ direction. The solution $H_2$ has no $y$ dependence, so the equation \eqref{ydu} gives trivial $y$ dependence on $u$, which has nothing to do with the gravitational wave. However, in this case, the $r$ dependence on $u$ can be solved independently by \eqref{rdu}. We consider the change in velocity of the particle in the $\gamma_1$ direction, that is the change in $\dot{r}(u)$ under influence of \eqref{pulse}. The $r$'s coordinate equation now takes the form,
\be\label{d2rdu2u>0}
\frac{d^2r}{du^2}=\frac{1}{k^2\ell^2r^3}-\frac{3}{4}Ar^2e^{bu}\,,
\ee
for $u<0$, and we should replace $b$ with $-b$ in \eqref{d2rdu2u>0} when $u>0$. To solve this equation, we assume that the gravitational impulsive wave is with small amplitude, so that the full solution can be expressed in a series expansion,
\be
r(u)=r_0(u)-\frac{3}{4}Ar_1(u)+\mathcal{O}(A^2)\,.
\ee
At zero order, we have a simple equation,
\be\label{r0eq}
\frac{d^2r_0}{du^2}=\frac{1}{k^2\ell^2r_0^3}\,,
\ee
with solution,
\be\label{r0}
r_0(u)=\sqrt{u^2+1/(k^2\ell^2)}\,,
\ee
comes from the contribution of the negative cosmological constant. $r_1(u)$ satisfies the first order equation,
\be\label{r1eq}
\frac{d^2r_1}{du^2}=\frac{-3}{k^2\ell^2r_0^4}r_1+r_0^2e^{bu}\,,
\ee
for $u<0$ and $r_0$ is given by \eqref{r0}. The solution of this equation is,
\bea\label{u<0}
r_1(u)&=&\frac{\sqrt{u^2+1/(k^2\ell^2)}}{4/(k\ell)}\left(\frac{u-i/(k\ell)}{u+i/(k\ell)}\right)(-i)\int_{u_0}^ue^{bs}(s+i/(k\ell))^2\sqrt{s^2+1/(k^2\ell^2)}ds\nonumber\\
&+&\text{hermitian conjugate}\,.
\eea
We have choose $u_0<0$ so that $r_1(u_0)=0$ and $\frac{dr_1}{du}|_{u=u_0}=0$. Now let us consider the change in $\frac{dr_1}{du}$ between $u=u_0$ and after the impulsive wave \eqref{pulse} passed over at $u=+\infty$. Since $u_0<0$, this is the time duration starts before the gravitational wave arrive and ends long after it had passed. The $u>0$ part of the solution $r_1(u)$ can be obtained by changing the sign of $b$ in \eqref{u<0} and connecting them at $u=0$,
\bea
r_1(u)&=&\frac{\sqrt{u^2+1/(k^2\ell^2)}}{4/(k\ell)}\left(\frac{u-i/(k\ell)}{u+i/(k\ell)}\right)\left(C_0-i\int_0^ue^{-bs}(s+i/(k\ell))^2\sqrt{s^2+1/(k^2\ell^2)}ds\right)\nonumber\\
&+&\text{hermitian conjugate}\,,
\eea
where
\be
C_0=-i\int_{u_0}^0e^{bs}(s+i/(k\ell))^2\sqrt{s^2+1/(k^2\ell^2)}ds\,.
\ee
The asymptotic behaviour of $r_1(u)$ at $u\to+\infty$ takes the form,
\be
r_1(u)|_{u\to+\infty}=\frac{u}{2/(k\ell)}\mr{Re}\left(C_0-i\int_0^ue^{-bs}(s+i/(k\ell))^2\sqrt{s^2+1/(k^2\ell^2)}ds\right)|_{u\to+\infty}\,.
\ee
So we have,
\be
\frac{dr_1}{du}|_{u\to+\infty}=\frac{k\ell}{2}\mr{Re}\left(C_0-i\int_0^{+\infty}e^{-bs}(s+i/(k\ell))^2\sqrt{s^2+1/(k^2\ell^2)}ds\right)\,.
\ee
Up to first order, the change in velocity in $\gamma_1$ direction after the gravitational impulsive wave had passed can be found,
\be
\Delta\left(\frac{dr}{du}\right)=\frac{\sqrt{u_0^2+1/(k^2\ell^2)}-u_0}{\sqrt{u_0^2+1/(k^2\ell^2)}}-\frac{3k\ell A}{8}\mr{Re}\left(C_0-i\int_0^{+\infty}e^{-bs}(s+i/(k\ell))^2\sqrt{s^2+1/(k^2\ell^2)}ds\right)\,,
\ee
where the particle is initially placed at $r(u_0)=\sqrt{u_0^2+1/(k^2\ell^2)}$ with velocity,
\be
\frac{dr}{du}|_{u=u_0}=\frac{u_0}{\sqrt{u_0^2+1/(k^2\ell^2)}}\,.
\ee
For $u_0\to0$, that is we initially put the particle at $r_0=\frac{1}{k\ell}$ with zero velocity, the change of velocity due to the gravitational wave can be written as
\be
\Delta\left(\frac{dr}{du}\right)_{H_2}=-\frac{3}{4}A\int_0^{+\infty}e^{-bs}s\sqrt{s^2+1/(k^2\ell^2)}ds\,.
\ee

Secondly, we consider the velocity memory of gravitational wave with circular periodic polarization. We take the solution
\be
H_3(u, r, y)=A\left(\frac{\cos\theta}{3}r^3+\frac{\sin\theta}{6}y^3+\frac{\sin\theta}{2}r^2y\right)
\ee
as in \eqref{h3}, where $A$ is the wave amplitude and $\theta$ is a function of $u$. In this case, the $r$ and $y$ components' equations take the form,
\bea
\frac{d^2r}{du^2}&=&\frac{1}{k^2\ell^2r^3}-\frac{A}{4}(\cos\theta r^2+\sin\theta ry)\,,\\
\frac{d^2y}{du^2}&=&-\frac{A}{8}\sin\theta(r^2+y^2)\,.
\eea
The way to solve the above equations is to do the perturbative expansion in terms of the wave amplitude $A$,
\bea
r(u)&=&r_0(u)-\frac{A}{4}r_1(u)+\mathcal{O}(A^2)\,,\\
y(u)&=&y_0(u)-\frac{A}{4}y_1(u)+\mathcal{O}(A^2)\,.
\eea
With these expansions, the first order of $r$ and $y$ components' equations are inhomogeneous linear differential equations,
\bea
&&\frac{d^2r_1}{du^2}=\frac{-3}{k^2\ell^2r_0^4}r_1+(\cos\theta r_0^2+\sin\theta r_0y_0),~~~~~~\frac{d^2r_0}{du^2}=\frac{1}{k^2\ell^2r_0^3}\,,\\
&&\frac{d^2y_1}{du^2}=\frac{\sin\theta}{2}(r_0^2+y_0^2),~~~~~~~~~~~~~~~~~~~~~~~~~~~~~~~~~~~~~\frac{d^2y_0}{du^2}=0\,.\label{y0y1}
\eea
The explicit solution of $r_1(u)$ in this case is,
\bea
r_1(u)&=&\frac{\sqrt{u^2+1/(k^2\ell^2)}}{4/(k\ell)}\left(\frac{u-i/(k\ell)}{u+i/(k\ell)}\right)(-i)\int_{0}^u\cos\theta(s+i/(k\ell))^2\sqrt{s^2+1/(k^2\ell^2)}ds\nonumber\\
&+&\text{hermitian conjugate}\,,
\eea
where we choose $r_0=\sqrt{u^2+1/(k^2\ell^2)}$ which results from the cosmological constant and $y_0=0$ for simplicity. We take the free falling particle initially at rest when $u=0$, that is $\frac{dr_0}{du}|_{u=0}=\frac{dr_1}{du}|_{u=0}=0$. At very late time, that is $u$ is much greater than the particle's inverse energy $1/(k\ell)$, we can write down the change of $r$ component velocity due to the gravitational wave up to first order,
\be
\Delta\left(\frac{dr}{du}\right)_{H_3}=-\frac{A}{4}\int_0^u\cos(\theta(s))s\sqrt{s^2+1/(k^2\ell^2)}ds\,.
\ee
From \eqref{y0y1}, the change of $y$ component velocity due to first order gravitaional wave can also be written as an integral,
\be
\Delta\left(\frac{dy}{du}\right)_{H_3}=-\frac{A}{8}\int_0^u\sin(\theta(s))(s^2+1/(k^2\ell^2))ds\,.
\ee
Given an explicit expression of $\theta(u)$, one can figure out the above integrals at any late time $u$. If we view the integrands above as an effective force acting on the particle in the $\gamma_1$ and $\gamma_2$ directions, we can see the periodic acceleration back and forth in each direction with an amplifier that is related to the background cosmological constant.

\appendix
\section{Linear functional derivatives}
Given a linear function $\mr{h}(a)$ and the fixed frame $\{e_i\}$, define its scalar coefficients as
\be
h_{ij}=e_i\cd\mr{h}(e_j)\,.
\ee
The partial derivative with respect to $\mr{h}(a)$ can be written interms of its coefficients,
\be
\p_{\mr{h}(a)}=a\cd e_je_i\p_{\mr{h}_{ij}}\,.
\ee
One can find,
\bea
\p_{\mr{h}(a)}(\mr{h}(b)\cd c)&=&a\cd e_je_i\p_{\mr{h}_{ij}}(\mr{h}_{mn}b^mc^n)\nonumber\\
&=&a\cd e_je_ic^ib^j\nonumber\\
&=&a\cd bc\,,
\eea
and
\bea
\p_{\mr{h}(a)}\la\mr{h}(b\w c)B\ra&=&\p_{\mr{h}(a)}\left\la\frac{1}{2}\mr{h}(b)\mr{h}(c)B-\frac{1}{2}\mr{h}(c)\mr{h}(b)B\right\ra\nonumber\\
&=&\dot{\p}_{\mr{h}(a)}\la\dot{\mr{h}}(b)\mr{h}(c)B\ra-\dot{\p}_{\mr{h}(a)}\la\dot{\mr{h}}(c)\mr{h}(b)B\ra\nonumber\\
&=&a\cd b\mr{h}(c)\cd B-a\cd c\mr{h}(b)\cd B\nonumber\\
&=&\mr{h}(a\cd(b\w c))\cd B\,.
\eea
where $B$ is an arbitrary bivector. The above result can be generalised to
\bea
\p_{\mr{h}(a)}\la\mr{h}(A_r)B_r\ra&=&\la\mr{h}(a\cd A_r)B_r\ra_1\,,\label{phrr}\\
\p_{\mr{h}(a)}\la\mr{h}(A)B\ra&=&\sum_r\la\mr{h}(a\cd A_r)B_r\ra_1\,,
\eea
where $A_r$ and $B_r$ are grade-r multivectors. In the second equation, the sum covers all possible grade-r terms in multivectors $A$ and $B$. By using \eqref{phrr} and \eqref{inverseF}, one can immediately derive,
\bea
\p_{\mr{h}(a)}\mr{det}(\mr{h})&=&\p_{\mr{h}(a)}\la\mr{h}(E_n)E_n^{-1}\ra=\la\mr{h}(a\cd E_n)E_n^{-1}\ra_1\nonumber\\
&=&\la\mr{h}(aE_n)E_n^{-1}\ra_1=\la E_n\mr{h}(E_n^{-1}a)\ra_1\nonumber\\
&=&\mr{det}(\mr{h})\bar{\mr{h}}^{-1}(a)\,.
\eea
The derivative of $\bar{\mr{h}}(b)$ with respect to $\mr{h}(a)$ is simply,
\bea
\p_{\mr{h}(a)}\bar{\mr{h}}(b)&=&\p_{\mr{h}(a)}(e^ie_i\cd\bar{\mr{h}}(b))\nonumber\\
&=&\p_{\mr{h}(a)}\la\mr{h}(e_i)b\ra e^i\nonumber\\
&=&(a\cd e_i)be^i\nonumber\\
&=&ba\,.
\eea
Similarly, one can define the derivative with respect to a bivector valued linear function $\Omega(a)$. Some of the useful results are,
\bea
\p_{\Omega(a)}\la\Omega(b)M\ra&=&a\cd b\la M\ra_2\,,\\
\p_{\Omega(b), a}\la c\cd\nabla\Omega(d)M\ra&=&a\cd cb\cd d\la M\ra_2\,,
\eea
where $M$ is an arbitrary multivector.

\section*{Acknowledgement}
I am grateful to Li-Ming Cao, Jiang Long, Bo Ning, and Zhao-Long Wang for for helpful discussions during the first ``National Symposium on Field theory and String Theory, 2020''. This work is supported by NSFC grant No. 12105045.

%===================================%
%<<<<<<<<<<<<< REFERENCES >>>>>>>>>>>>>%
%======================================%

\end{document}